\documentclass[aps,prl,preprint,superscriptaddress,showpacs]{revtex4-1}
\usepackage{graphics} 
\usepackage{amsmath} 
\usepackage{graphicx} 
\usepackage{dcolumn} 
\usepackage{bm} 
\usepackage{color}
\usepackage{natbib}

\newcommand{\eps}{\varepsilon}
\newcommand{\ro}{\rho}

\newcommand{\spu}{\hat{\sigma}^{+}_{1}}

\newcommand{\smd}{\hat{\sigma}^{-}_{2}}

\newcommand{\w}{\omega}
\newcommand{\wt}{\omega(t)}
\newcommand{\wo}{\omega_{0}}
\newcommand{\epst}{\tilde{\varepsilon}}

\newcommand{\ud}{ |\uparrow \downarrow\rangle}
\newcommand{\du}{|\downarrow \uparrow\rangle}

\newcommand{\pvec}{|\Psi\rangle}

\begin{document}

\title{Creation of entangled states in coupled quantum dots via adiabatic rapid passage}
\author{C. Creatore} 
\author{R. T. Brierley}
\author{R. T. Phillips}
\affiliation{Cavendish Laboratory, University of Cambridge, CB3 0HE, Cambridge, United Kingdom}
\author{P. B. Littlewood}
\affiliation{Cavendish Laboratory, University of Cambridge, CB3 0HE, Cambridge, United Kingdom}
\affiliation{Argonne National Laboratory, Argonne IL 60439, USA}
\affiliation{James Franck Inst., University of Chicago, Chicago IL 60637}
\author{P. R. Eastham}
\affiliation{School of Physics, Trinity College, Dublin 2, Ireland}
\begin{abstract}
  Quantum state preparation through external control is fundamental to
  established methods in quantum information processing and in studies
  of dynamics.
  In this respect, excitons in semiconductor quantum dots (QDs) are of
  particular interest since their coupling to light allows them to be
  driven into a specified state using the coherent interaction with a
  tuned optical field such as an external laser
  pulse. 
  We propose a protocol, based on adiabatic rapid passage, for the
  creation of entangled states in an ensemble of pairwise coupled
  two-level systems, such as an ensemble of QD molecules. We show by
  quantitative analysis using realistic parameters for semiconductor
  QDs that this method is feasible where other approaches are
  unavailable.  Furthermore, this scheme can be generically
  transferred to some other physical systems including circuit QED,
  nuclear and electron spins in solid-state environments, and photonic
  coupled cavities.
\end{abstract}
\date{\today}
\maketitle

Whether one is interested in methods of quantum optics or computation,
or more general many-body dynamics, it is important to be able to
prepare interacting few-level systems in initial states from pure
product states to more complex entanglements. Excitons in
low-dimensional semiconductors exemplify such a level structure.  They
are of interest both because their strong coupling to light provides
communication with the external world, and because photon-mediated
interactions between excitons can be substantial.  However, the lack
of degeneracy due to the (accidental or deliberate) inhomogeneous
spread of the energy levels becomes a challenge for addressing
different levels -- either independently, for example to invert a
single one, or in pairs, for entangling a specified set. Control of
the population of excitonic states in quantum dots (QDs) has been
demonstrated using resonant transform-limited light fields such as
$\pi$ Rabi
pulses~\cite{Stievater2001,Kamada2001,Htoon2002,Zrenner2002}.
Although effective in certain cases, Rabi flopping has the
disadvantage for population transfer that the final state is sensitive
to variations in dipole couplings, dot energy and disorder. Such
disorder becomes a serious challenge if one wishes to scale this
approach beyond a single quantum dot, and similar limitations arise in
many other systems. More recently, population transfer has been
achieved~\cite{Wu2011,Simon2011} using a protocol called adiabatic
rapid passage (ARP)~\cite{Malinovsky2001}.  Within this scheme the
frequency of the laser pulse is swept through that of the target state
and, if whole process is performed adiabatically, this target state
will be populated with high efficiency. Unlike the Rabi approach, ARP
is largely unaffected by all the variations previously mentioned, and
this has prompted theoretical proposals for its use in preparing
populations in ensembles of QDs~\cite{Schmidgall2010}, as well as
proposals to realize a non-equilibrium Bose-Einstein
condensate~\cite{Eastham2009,Schmidgall2010,Brierley2010} and quantum
operations
~\cite{Unanyan2001,Unanyan2002,Paspalakis2004,Fabian2005,Troiani2006,Saikin2008,Gauger2008}.

The aim of this paper is to show how ARP may be generalized to create
entangled states in a disordered system, comprising a set of two-level
systems coupled together in pairs. Such an ensemble could be realized
using quantum dot molecules, with a possible intra-molecular coupling
mechanism being resonant F$\rm\ddot{o}$rster
transfer~\cite{Lovett2003,Lovett2005,Loss2006,Paspalakis2004}. Implementations
would also be feasible in other systems such as photonic coupled
cavities with photons tunneling between neighboring cavities and
having onsite interaction~\cite{Hartmann2008}, in arrays of
superconducting qubits with an exchange coupling mediated by virtual
photons~\cite{Wallraff2007,Wallraff2011}, or using electron spin
states in coupled semiconductor impurities~\cite{Morley2010}. 

The important feature of our proposal is that it requires neither
precise control, nor reproducibility, of the pairs. Instead we exploit
the disorder to allow those pairs with entangled states to be
spectrally identified within an ensemble. The robustness of ARP then
allows a pulse to be constructed which drives those pairs into their
entangled states, without creating other excitations. Thus, as shown
in Fig.~\ref{fig:figure2}, the entanglement of formation per
excitation is very close to one, even in a strongly disordered
system. Our approach represents a significant simplification and
improvement of the protocol required for the production of entangled
states in realistic systems, since previous proposals have relied on
coupling two states through a further level~\cite{Troiani2006}, or
consider only a single and fine-tuned
system~\cite{Quiroga99,Unanyan2001,Unanyan2002,Paspalakis2004,Obada2007}.

To simplify the notation we consider the limit in which the states of
the individual quantum dots form two-level systems that may be
represented in terms of a Bloch vector or pseudospin: each dot either
contains no exciton (spin down) or a single exciton (spin up). The
generalization to allow for multiple states on the dot, for example
due to the exciton spin structure, is straightforward. The relevant
Hamiltonian for two coupled dots, in a frame rotating with the
time-dependent frequency $\wt$ of the ARP pulse and in the rotating
wave approximation, is then ($\hbar=1$):
\begin{equation}
\label{eq:H2}
\hat{H}=\sum_{i=1,2}\left[\frac{\epst_i}{2}\hat{\sigma}_{i}^{z} + g(t)(\hat{\sigma}_{i}^{+} + \hat{\sigma}_{i}^{-}) \right]- j_{T}(\spu\smd + \smd\spu)\,,
\end{equation}
where $\epst_{i}=\eps_{i} - \w(t)$, $\eps_{i}$ is the transition
energy for dot $i=1,2$, $g(t)$ the amplitude of the driving field used
to perform ARP, $\sigma_{i}$ the Pauli operators describing the state
of dot $i$, and $j_{T}$ an exchange interaction between the two sites
($\hbar=1$). The ARP pulse can be decomposed into its amplitude and
frequency, i.e., $F(t)=g(t){\rm exp}\left(i\int
  \omega(t^{\prime})dt^{\prime}\right)$. For definiteness we consider
a Gaussian linearly chirped pulse:
\begin{equation}
\label{eq:pulse}
F(t)=g(t)e^{i\omega(t)t}=g_{0}e^{-t^{2}/2\tau^{2}}e^{i\omega(t)t}\,,\,\,\, \w(t)=\wo + \alpha t\,,
\end{equation} 
with $g_{0}$ the pulse amplitude, $\tau$ the temporal width of the
pulse, $\w_{0}$ its central frequency, and $\alpha$ the chirp. We take
the pulse width to define units of time and energy, and use dimensionless
parameters $\alpha\tau^{2}$, $g_{0}\tau$ and $j_{T}\tau$.

In general, ARP schemes are generalizations of the Landau-Zener
problem~\cite{Landau,Zener} to a time-dependent and controllable
mixing $g(t)$ between the levels. Consider first noninteracting dots
($j_T=0$): if the driving frequency sweeps through the transition
frequency of a dot, $\epst_{i}$ passes through zero, and the lowest
energy eigenstate changes from an empty site to being fully
occupied. In the Landau-Zener problem there is a constant
hybridization $g_0$ between these levels to create an anticrossing,
and a system in the initial ground state will be driven into the
excited state with probability $P_{\rm inv}=1-{\rm
  exp}(-g_0^{2}/\alpha)$. Thus as the adiabaticity is increased, by
reducing the chirp $\alpha$ or increasing the hybridization $g_0$, the
final population rapidly approaches one. In the case of ARP, the
coupling term $g(t)$ is not constant and, in order for the transition
to occur, the two levels of the system need to be coupled long enough
so that the character of the eigenstates changes smoothly, i.e.,
adiabatically. This gives the criterion
$\alpha\tau^2\gg1$~\cite{Malinovsky2001}.  Furthermore, the ARP scheme
can be made spectrally selective. One can choose to excite either one
dot or the other by timing the center of the optical pulse to occur
when the chirp has driven a particular level to zero, and ensuring
that the pulse is negligible at the other crossing. The available
discrimination in the energy levels is thus of order $\hbar/\tau$.

This discrimination could be used to create entangled states in the
ideal case of a pair of identical two-level
systems~\cite{Unanyan2001,Unanyan2002,Paspalakis2004}. In this case
the Hamiltonian (\ref{eq:H2}) is, in the absence of the driving field,
diagonal in the singlet/triplet basis
$\{|S\rangle,|T_{-}\rangle,|T_{0}\rangle,|T_{+}\rangle\}$, with
$|S\rangle=(1/\sqrt{2})\left(|\downarrow\uparrow\rangle -
|\uparrow\downarrow\rangle\right)$,
$|T_{-}\rangle=|\downarrow\downarrow\rangle$,
$|T_{0}\rangle=(1/\sqrt{2})\left(|\downarrow\uparrow\rangle +
|\uparrow\downarrow\rangle\right)$,
$|T_{+}\rangle=|\uparrow\uparrow\rangle$. The intermediate states
$|T_{0}\rangle$ and $|S\rangle$ between the ground state and the fully
occupied state $|T_{+}\rangle$ are spatially entangled if the coupling
dominates over the detuning, $\Delta\eps=\eps_{1}-\eps_{2}\lesssim
j_{T}$. For a single pair of identical dots
($\eps_{1}=\eps_{2}=\eps_{0}$), the use of ARP to create this
entangled state is illustrated in Fig. 1(a) which shows the time
evolution of the eigenvalues of the Hamiltonian Eq. (1) when the
levels are coupled by a linearly chirped Gaussian pump ($g_{0}\tau=5$,
colored continuous lines) and uncoupled ($g_{0}\tau=0$, thin dashed
lines). By chirping through the level crossing (A) between the states
$|T_{-}\rangle$ and $|T_{0}\rangle$ (at $\wo=\eps_{0} - j_{T}$) with a
pulse duration such that coupling switches off before the
$|T_{0}\rangle\rightarrow|T_{+}\rangle$ crossing (B), the system will
be adiabatically driven and left in the entangled triplet state
$|T_{0}\rangle$.

We now consider the parameters required for this procedure for a
single degenerate pair. In this case the population of the two-dot
system in the desired final state is equally shared between
$|\uparrow\downarrow\rangle$ and $|\downarrow\uparrow\rangle$, so
$\ro_{\uparrow\downarrow}=\ro_{\downarrow\uparrow}=0.5$.  The element
of the density matrix $\ro_{\uparrow\downarrow}$
($=\ro_{\downarrow\uparrow}$), can thus be used as a measure of the
insensitivity of the entanglement to variation of inter-dot coupling
and chirp.  This is illustrated in Fig. 1(b) which shows
$\ro_{\uparrow\downarrow}$, obtained by numerical solution of the
Liouville equation $d\ro/dt=-i\left[H,\ro\right]$, as a function of
the dimensionless parameters $\alpha\tau^{2}$ and $j_{T}\tau$.  The
region where $\ro_{\uparrow\downarrow}=0.5$ is seen to extend over a
large range of values of chirp and exchange coupling. 
\begin{figure}[t!]
\label{fig:figure1}
\begin{center}
\includegraphics[width=0.60\textwidth]{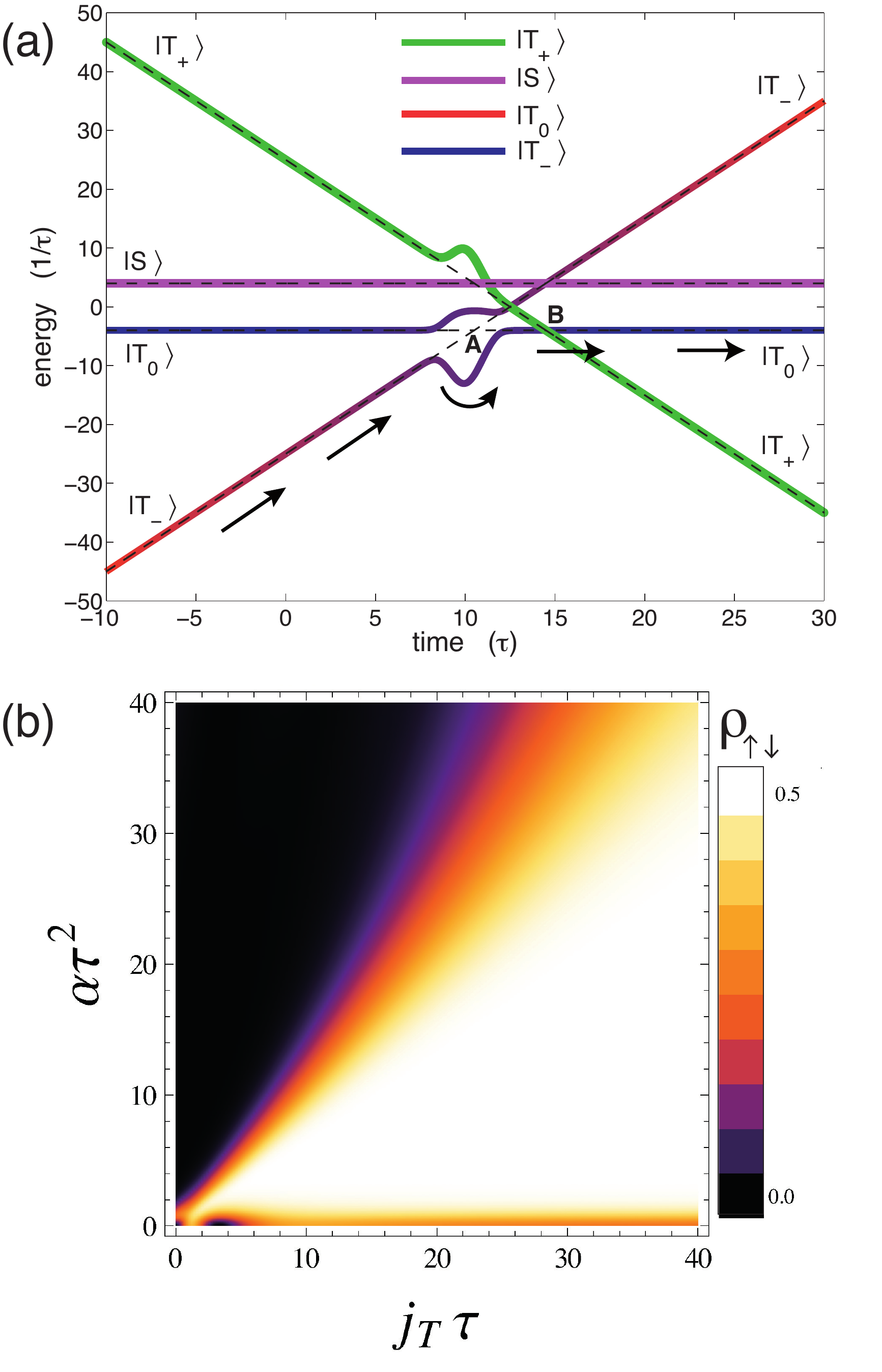}\\
\end{center}
\caption{(Color online) (a) Time-dependent energy levels of a coupled
  system of two identical dots. The continuous colored lines indicate
  the energy levels with a Gaussian driving pulse ($g_{0}\tau=5$) of
  duration $\tau$, while the thin dashed lines are the undriven levels
  ($g_{0}=0$). The chirp and interdot couplings are $\alpha\tau^{2}=2$
  and $j_{T}\tau=4$. The central frequency of the ARP pulse is
  $\eps_{0} - j_{T}$, resonant with the transition between the ground
  state $|T_{-}\rangle$ and the entangled state $|T_{0}\rangle$ (see
  point A).  (b) Final density matrix element
  $\ro_{\uparrow\downarrow}$, calculated as a function of the
  dimensionless linear chirp $\alpha\tau^{2}$ and exchange coupling
  $j_{T}\tau$. The white region ($\ro_{\uparrow\downarrow}=0.5$) shows
  the range of values of chirp and exchange coupling where the pair is
  driven into the entangled state $|T_{0}\rangle$ with high probability.}
\end{figure}
The required parameters are fully compatible with a linearly chirped pulse similar
to that used to invert a single semiconductor QD~\cite{Wu2011}: in
that case we have considered a transform-limited pulse width of 2 ps
and a chirped temporal width $\tau=4.5$ ps. Thus for
$\alpha\tau^2\approx 2$ we require a dimensionless coupling
$j_{T}\tau\gtrsim 4$ [see Fig. 1] for the ARP transition to the
entangled state to occur, which corresponds to a value of F$\rm
\ddot{o}$rster coupling of 0.6 meV. Previous studies have estimated an
upper limit of 10 meV for F$\rm \ddot{o}$rster coupling in
semiconductor quantum dots~\cite{Loss2006}. Thus the scheme could be
implemented using two stacked (vertical) QDs at a distance of few
nanometers, coupled by F$\rm\ddot{o}$rster energy transfer, but with
no single-particle tunneling. Such conditions can be achieved in
InAs/GaAs coupled QDs~\cite{YamauchiAPL05,KrennerPRL05}. The scheme
would also apply with single-particle tunneling or indirect excitons,
provided only that target entangled states can be identified.

We now show how this adiabatic protocol can be generalized to generate
entanglement in realistic non-degenerate systems such as ensembles of
QDs. In this case the entanglement pulse can be spectrally tuned to
address a specific pair of dots, and several pairs can be entangled
within an ensemble by superposing such pulses. 

It is important to recognize that the requirement of exact degeneracy
of the uncoupled transition is relaxed up to the magnitude of the
coupling energy. This affords a route to practical realizations of the
scheme, as the coupling energy and level splitting can be traded to
optimize the probability of producing an appropriate double-dot
structure. In the following we consider an ensemble of such systems,
modeled here by an average coupling strength and having an
inhomogeneous distribution of energies. In each coupled pair the
energy levels are not degenerate, and only pairs with detunings
smaller than the interdot coupling strength can be entangled; these
specific pairs must be identified spectroscopically prior to ARP
manipulation. Spectrally-selected components of a broad-band pulse,
each close to resonance with a particular chosen pair, can then be
chirped in the same linear optical process, such as a grating-based
delay stage.  We test this scheme simulating an ensemble consisting of
thirty pairs of dots with energies taken from a Gaussian distribution
of standard deviation 10 meV, and coupled by an average coupling
strength $j_{T}=$2 meV.  In one typical realization taken as an
example, three couples can be entangled as their energy levels are
detuned by an amount smaller than $j_{T}/2$. To evaluate the
entanglement produced, we use the {\it entanglement of formation}
(EOF), which is a widely accepted measure of the entanglement for
bipartite states. For pure states it is the entropy~\cite{Wootters97}:
\begin{equation}
\label{eq:measure}
E(\pvec)=S(\rho)\,,
\end{equation}
where $S(\rho)=-{\rm Tr}\left(\rho^{\rm red}{\rm log_{2}}\rho^{\rm
  red}\right)$ is the von Neumann entropy and $\rho^{\rm red}={\rm
  Tr^{red}|\Psi\rangle\langle\Psi|}$ is the reduced density matrix
obtained by tracing the whole system density matrix
$|\Psi\rangle\langle\Psi|$ over one of the two subsystems of which the
pure state $\pvec$ consists.  In this case the two subsystems are the
two paired dots. The EOF (or entropy) ranges from zero (for a product
state) to ${\rm log_{2}}N$ for a maximally entangled state of two
$N$-state particles. Hence, the EOF of the triplet state
$|T_{0}\rangle=(1/\sqrt{2})(\ud+\du)$ (which is a maximally entangled
Bell state) is equal to 1. Furthermore, the properties of the EOF as a
natural measure of the entanglement include that the entanglement of
independent systems, such as pairs of coupled dots in an ensemble, is
additive.

Figure 2 shows the EOF and the total excitation in our example
realization calculated as a function of the strength of the applied
chirped Gaussian pulse.  The excitation induced in the system after
the application of an external pulse has been evaluated as $X=\rm
Tr(\rho\hat{X})$, the excitation operator $\rm \hat{X}$ given by
$\rm{\hat{X}}=(\sigma_{1}^{z}+1)/2 + (\sigma_{2}^{z}+1)/2$ and
$\sigma_{1,2}^{z}$ being the two-particles spin operators for dot 1
and 2. We have used $\tau=15$ ps and $\alpha=0.01$ ps$^{-2}$
($\alpha\tau^{2}$=2.25) yielding a small value of $\alpha\tau$ --
which determines the energy range spanned by each component of the ARP
pulse -- while keeping the whole process in the adiabatic regime
$\alpha\tau^{2}>1$, so that few states apart from the entangled ones
will be excited. The total entanglement (dashed red curve) always
deviates from the ideal value of 3, i.e., the total number of triplet
states, due to the unavoidable excitation of other states which are
not entangled, but yield a small amount of entropy. Furthermore, as
the intensity of the pulse increases, other level crossings become
adiabatic and other states such as the full inversion
$|\uparrow\uparrow\rangle$ can be populated, thus explaining the
behavior of the excitation (continuous black) curve. However, the EOF
normalized with respect to the total excitation (dashed-dotted blue
curve), which is an effective measure of the total entanglement
produced in the system, is close to the ideal value of 1 ($\approx$
0.85 for all the different pump strengths considered). In order to
validate these results we have simulated 50 different realizations of
Gaussian-distributed coupled dots, and found an average value for the
normalized EOF always $\approx$ 0.8 for several realistic values of
pulse strengths.
\begin{figure}[t!]
\begin{center}
\includegraphics[width=0.60\textwidth]{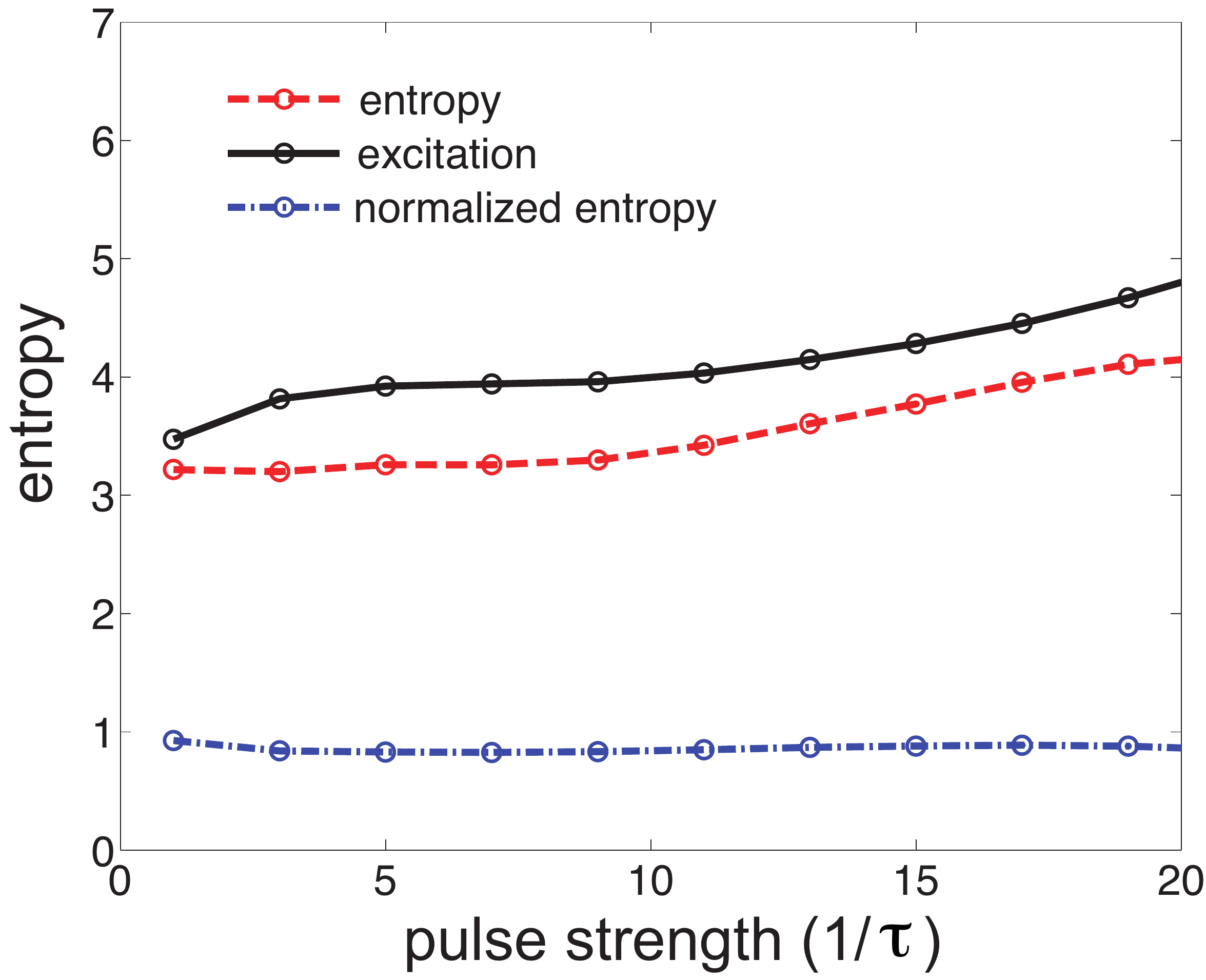}\\
\end{center}
\caption{(Color online) Predicted entanglement generated when an
  ensemble of 30 coupled pairs of dots is driven by an ARP pulse. The
  pulse is constructed to drive the 3 strongly-coupled pairs in the
  ensemble into their entangled state (see text). The red dashed
  curve shows the total entanglement entropy (or entropy of
  formation), the black solid curve the total excitation, and the
  blue dot-dashed curve the entanglement entropy normalized by the
  total excitation. The horizontal axis is the strength of the driving
  field, defined as twice the peak Rabi splitting in a single dot,
    given in units of $1/\tau$ ($\hbar=1$).}
\label{fig:figure2}
\end{figure}

In conclusion, we have shown the benefits of ARP for the generation
of entangled stated in ensembles of coupled quantum systems such as
inhomogeneously distributed QD excitons. Our calculations, based on
realistic values for F$\rm \ddot{o}$rster coupling and dot
distribution, provide a feasible route for the realization of
entanglement in solid state systems, which is more practical than
other approaches due to the flexible and tunable parameters of ARP. In
principle, the same methodology can apply to microcavity Josephson
junctions, to coupled electron spins in semiconductors (using ESR),
and to coupled nuclear spins (via NMR). The latter arena is the
genesis of the ARP technique, though we are not aware of it being
explicitly used in the form proposed here.

C.C. and R.T.P. acknowledge support from EPSRC under Grant EP/F040075/1. 
P.R.E. acknowledges support from Science Foundation Ireland under Grant 09/SIRG/I1592. 
P.B.L. acknowledges support by DOE under FWP 70069.


\begin{thebibliography}{31}%
\makeatletter
\providecommand \@ifxundefined [1]{%
 \@ifx{#1\undefined}
}%
\providecommand \@ifnum [1]{%
 \ifnum #1\expandafter \@firstoftwo
 \else \expandafter \@secondoftwo
 \fi
}%
\providecommand \@ifx [1]{%
 \ifx #1\expandafter \@firstoftwo
 \else \expandafter \@secondoftwo
 \fi
}%
\providecommand \natexlab [1]{#1}%
\providecommand \enquote  [1]{``#1''}%
\providecommand \bibnamefont  [1]{#1}%
\providecommand \bibfnamefont [1]{#1}%
\providecommand \citenamefont [1]{#1}%
\providecommand \href@noop [0]{\@secondoftwo}%
\providecommand \href [0]{\begingroup \@sanitize@url \@href}%
\providecommand \@href[1]{\@@startlink{#1}\@@href}%
\providecommand \@@href[1]{\endgroup#1\@@endlink}%
\providecommand \@sanitize@url [0]{\catcode `\\12\catcode `\$12\catcode
  `\&12\catcode `\#12\catcode `\^12\catcode `\_12\catcode `\%12\relax}%
\providecommand \@@startlink[1]{}%
\providecommand \@@endlink[0]{}%
\providecommand \url  [0]{\begingroup\@sanitize@url \@url }%
\providecommand \@url [1]{\endgroup\@href {#1}{\urlprefix }}%
\providecommand \urlprefix  [0]{URL }%
\providecommand \Eprint [0]{\href }%
\providecommand \doibase [0]{http://dx.doi.org/}%
\providecommand \selectlanguage [0]{\@gobble}%
\providecommand \bibinfo  [0]{\@secondoftwo}%
\providecommand \bibfield  [0]{\@secondoftwo}%
\providecommand \translation [1]{[#1]}%
\providecommand \BibitemOpen [0]{}%
\providecommand \bibitemStop [0]{}%
\providecommand \bibitemNoStop [0]{.\EOS\space}%
\providecommand \EOS [0]{\spacefactor3000\relax}%
\providecommand \BibitemShut  [1]{\csname bibitem#1\endcsname}%
\let\auto@bib@innerbib\@empty
\bibitem [{\citenamefont {Stievater}\ \emph {et~al.}(2001)\citenamefont
  {Stievater}, \citenamefont {Li}, \citenamefont {Steel}, \citenamefont
  {Gammon}, \citenamefont {Katzer}, \citenamefont {Park}, \citenamefont
  {Piermarocchi},\ and\ \citenamefont {Sham}}]{Stievater2001}%
  \BibitemOpen
  \bibfield  {author} {\bibinfo {author} {\bibfnamefont {T.~H.}\ \bibnamefont
  {Stievater}}, \bibinfo {author} {\bibfnamefont {X.}~\bibnamefont {Li}},
  \bibinfo {author} {\bibfnamefont {D.~G.}\ \bibnamefont {Steel}}, \bibinfo
  {author} {\bibfnamefont {D.}~\bibnamefont {Gammon}}, \bibinfo {author}
  {\bibfnamefont {D.~S.}\ \bibnamefont {Katzer}}, \bibinfo {author}
  {\bibfnamefont {D.}~\bibnamefont {Park}}, \bibinfo {author} {\bibfnamefont
  {C.}~\bibnamefont {Piermarocchi}}, \ and\ \bibinfo {author} {\bibfnamefont
  {L.~J.}\ \bibnamefont {Sham}},\ }\href@noop {} {\bibfield  {journal}
  {\bibinfo  {journal} {Phys. Rev. Lett.}\ }\textbf {\bibinfo {volume} {87}},\
  \bibinfo {pages} {133603} (\bibinfo {year} {2001})}\BibitemShut {NoStop}%
\bibitem [{\citenamefont {Kamada}\ \emph {et~al.}(2001)\citenamefont {Kamada},
  \citenamefont {Gotoh}, \citenamefont {Temmyo}, \citenamefont {Takagahara},\
  and\ \citenamefont {Ando}}]{Kamada2001}%
  \BibitemOpen
  \bibfield  {author} {\bibinfo {author} {\bibfnamefont {H.}~\bibnamefont
  {Kamada}}, \bibinfo {author} {\bibfnamefont {H.}~\bibnamefont {Gotoh}},
  \bibinfo {author} {\bibfnamefont {J.}~\bibnamefont {Temmyo}}, \bibinfo
  {author} {\bibfnamefont {T.}~\bibnamefont {Takagahara}}, \ and\ \bibinfo
  {author} {\bibfnamefont {H.}~\bibnamefont {Ando}},\ }\href@noop {} {\bibfield
   {journal} {\bibinfo  {journal} {Phys. Rev. Lett.}\ }\textbf {\bibinfo
  {volume} {87}},\ \bibinfo {pages} {246401} (\bibinfo {year}
  {2001})}\BibitemShut {NoStop}%
\bibitem [{\citenamefont {Htoon}\ \emph {et~al.}(2002)\citenamefont {Htoon},
  \citenamefont {Takagahara}, \citenamefont {Kulik}, \citenamefont {Baklenov},
  \citenamefont {Holmes},\ and\ \citenamefont {Shih}}]{Htoon2002}%
  \BibitemOpen
  \bibfield  {author} {\bibinfo {author} {\bibfnamefont {H.}~\bibnamefont
  {Htoon}}, \bibinfo {author} {\bibfnamefont {T.}~\bibnamefont {Takagahara}},
  \bibinfo {author} {\bibfnamefont {D.}~\bibnamefont {Kulik}}, \bibinfo
  {author} {\bibfnamefont {O.}~\bibnamefont {Baklenov}}, \bibinfo {author}
  {\bibfnamefont {A.~L.}\ \bibnamefont {Holmes}}, \ and\ \bibinfo {author}
  {\bibfnamefont {C.~K.}\ \bibnamefont {Shih}},\ }\href@noop {} {\bibfield
  {journal} {\bibinfo  {journal} {Phys. Rev. Lett.}\ }\textbf {\bibinfo
  {volume} {88}},\ \bibinfo {pages} {087401} (\bibinfo {year}
  {2002})}\BibitemShut {NoStop}%
\bibitem [{\citenamefont {Zrenner}\ \emph {et~al.}(2002)\citenamefont
  {Zrenner}, \citenamefont {Beham}, \citenamefont {Stufler},\ and\
  \citenamefont {Findeis}}]{Zrenner2002}%
  \BibitemOpen
  \bibfield  {author} {\bibinfo {author} {\bibfnamefont {A.}~\bibnamefont
  {Zrenner}}, \bibinfo {author} {\bibfnamefont {E.}~\bibnamefont {Beham}},
  \bibinfo {author} {\bibfnamefont {S.}~\bibnamefont {Stufler}}, \ and\
  \bibinfo {author} {\bibfnamefont {F.}~\bibnamefont {Findeis}},\ }\href@noop
  {} {\bibfield  {journal} {\bibinfo  {journal} {Nature}\ }\textbf {\bibinfo
  {volume} {418}},\ \bibinfo {pages} {612} (\bibinfo {year}
  {2002})}\BibitemShut {NoStop}%
\bibitem [{\citenamefont {Wu}\ \emph {et~al.}(2011)\citenamefont {Wu},
  \citenamefont {Piper}, \citenamefont {Ediger}, \citenamefont {Brereton},
  \citenamefont {Schmidgall}, \citenamefont {Eastham}, \citenamefont {Hugues},
  \citenamefont {Hopkinson},\ and\ \citenamefont {Phillips}}]{Wu2011}%
  \BibitemOpen
  \bibfield  {author} {\bibinfo {author} {\bibfnamefont {Y.}~\bibnamefont
  {Wu}}, \bibinfo {author} {\bibfnamefont {I.~M.}\ \bibnamefont {Piper}},
  \bibinfo {author} {\bibfnamefont {M.}~\bibnamefont {Ediger}}, \bibinfo
  {author} {\bibfnamefont {P.}~\bibnamefont {Brereton}}, \bibinfo {author}
  {\bibfnamefont {E.~R.}\ \bibnamefont {Schmidgall}}, \bibinfo {author}
  {\bibfnamefont {P.~R.}\ \bibnamefont {Eastham}}, \bibinfo {author}
  {\bibfnamefont {M.}~\bibnamefont {Hugues}}, \bibinfo {author} {\bibfnamefont
  {M.}~\bibnamefont {Hopkinson}}, \ and\ \bibinfo {author} {\bibfnamefont
  {R.~T.}\ \bibnamefont {Phillips}},\ }\href@noop {} {\bibfield  {journal}
  {\bibinfo  {journal} {Phys. Rev. Lett.}\ }\textbf {\bibinfo {volume} {106}},\
  \bibinfo {pages} {067401} (\bibinfo {year} {2011})}\BibitemShut {NoStop}%
\bibitem [{\citenamefont {Simon}\ \emph {et~al.}(2011)\citenamefont {Simon},
  \citenamefont {Belhadj}, \citenamefont {Chatel}, \citenamefont {Amand},
  \citenamefont {Renucci}, \citenamefont {Lemaitre}, \citenamefont {Krebs},
  \citenamefont {Dalgarno}, \citenamefont {Warburton}, \citenamefont {Marie},\
  and\ \citenamefont {Urbaszek}}]{Simon2011}%
  \BibitemOpen
  \bibfield  {author} {\bibinfo {author} {\bibfnamefont {C.-M.}\ \bibnamefont
  {Simon}}, \bibinfo {author} {\bibfnamefont {T.}~\bibnamefont {Belhadj}},
  \bibinfo {author} {\bibfnamefont {B.}~\bibnamefont {Chatel}}, \bibinfo
  {author} {\bibfnamefont {T.}~\bibnamefont {Amand}}, \bibinfo {author}
  {\bibfnamefont {P.}~\bibnamefont {Renucci}}, \bibinfo {author} {\bibfnamefont
  {A.}~\bibnamefont {Lemaitre}}, \bibinfo {author} {\bibfnamefont
  {O.}~\bibnamefont {Krebs}}, \bibinfo {author} {\bibfnamefont {P.~A.}\
  \bibnamefont {Dalgarno}}, \bibinfo {author} {\bibfnamefont {R.~J.}\
  \bibnamefont {Warburton}}, \bibinfo {author} {\bibfnamefont {X.}~\bibnamefont
  {Marie}}, \ and\ \bibinfo {author} {\bibfnamefont {B.}~\bibnamefont
  {Urbaszek}},\ }\href@noop {} {\bibfield  {journal} {\bibinfo  {journal}
  {Phys. Rev. Lett.}\ }\textbf {\bibinfo {volume} {106}},\ \bibinfo {pages}
  {166801} (\bibinfo {year} {2011})}\BibitemShut {NoStop}%
\bibitem [{\citenamefont {Malinovsky}\ and\ \citenamefont
  {Krause}(2001)}]{Malinovsky2001}%
  \BibitemOpen
  \bibfield  {author} {\bibinfo {author} {\bibfnamefont {V.}~\bibnamefont
  {Malinovsky}}\ and\ \bibinfo {author} {\bibfnamefont {J.}~\bibnamefont
  {Krause}},\ }\href@noop {} {\bibfield  {journal} {\bibinfo  {journal} {Eur.
  Phys. J. D}\ }\textbf {\bibinfo {volume} {14}},\ \bibinfo {pages} {147}
  (\bibinfo {year} {2001})}\BibitemShut {NoStop}%
\bibitem [{\citenamefont {Schmidgall}\ \emph {et~al.}(2010)\citenamefont
  {Schmidgall}, \citenamefont {Eastham},\ and\ \citenamefont
  {Phillips}}]{Schmidgall2010}%
  \BibitemOpen
  \bibfield  {author} {\bibinfo {author} {\bibfnamefont {E.~R.}\ \bibnamefont
  {Schmidgall}}, \bibinfo {author} {\bibfnamefont {P.~R.}\ \bibnamefont
  {Eastham}}, \ and\ \bibinfo {author} {\bibfnamefont {R.~T.}\ \bibnamefont
  {Phillips}},\ }\href@noop {} {\bibfield  {journal} {\bibinfo  {journal}
  {Phys. Rev. B}\ }\textbf {\bibinfo {volume} {81}},\ \bibinfo {pages} {195306}
  (\bibinfo {year} {2010})}\BibitemShut {NoStop}%
\bibitem [{\citenamefont {Eastham}\ and\ \citenamefont
  {Phillips}(2009)}]{Eastham2009}%
  \BibitemOpen
  \bibfield  {author} {\bibinfo {author} {\bibfnamefont {P.~R.}\ \bibnamefont
  {Eastham}}\ and\ \bibinfo {author} {\bibfnamefont {R.~T.}\ \bibnamefont
  {Phillips}},\ }\href@noop {} {\bibfield  {journal} {\bibinfo  {journal}
  {Phys. Rev. B}\ }\textbf {\bibinfo {volume} {79}},\ \bibinfo {pages} {165303}
  (\bibinfo {year} {2009})}\BibitemShut {NoStop}%
\bibitem [{\citenamefont {Brierley}\ and\ \citenamefont
  {Eastham}(2010)}]{Brierley2010}%
  \BibitemOpen
  \bibfield  {author} {\bibinfo {author} {\bibfnamefont {R.~T.}\ \bibnamefont
  {Brierley}}\ and\ \bibinfo {author} {\bibfnamefont {P.~R.}\ \bibnamefont
  {Eastham}},\ }\href@noop {} {\bibfield  {journal} {\bibinfo  {journal} {Phys.
  Rev. B}\ }\textbf {\bibinfo {volume} {82}},\ \bibinfo {pages} {035317}
  (\bibinfo {year} {2010})}\BibitemShut {NoStop}%
\bibitem [{\citenamefont {Unanyan}\ \emph {et~al.}(2001)\citenamefont
  {Unanyan}, \citenamefont {Vitanov},\ and\ \citenamefont
  {Bergmann}}]{Unanyan2001}%
  \BibitemOpen
  \bibfield  {author} {\bibinfo {author} {\bibfnamefont {R.~G.}\ \bibnamefont
  {Unanyan}}, \bibinfo {author} {\bibfnamefont {N.~V.}\ \bibnamefont
  {Vitanov}}, \ and\ \bibinfo {author} {\bibfnamefont {K.}~\bibnamefont
  {Bergmann}},\ }\href@noop {} {\bibfield  {journal} {\bibinfo  {journal}
  {Phys. Rev. Lett.}\ }\textbf {\bibinfo {volume} {87}},\ \bibinfo {pages}
  {137902} (\bibinfo {year} {2001})}\BibitemShut {NoStop}%
\bibitem [{\citenamefont {Unanyan}\ \emph {et~al.}(2002)\citenamefont
  {Unanyan}, \citenamefont {Fleischhauer}, \citenamefont {Vitanov},\ and\
  \citenamefont {Bergmann}}]{Unanyan2002}%
  \BibitemOpen
  \bibfield  {author} {\bibinfo {author} {\bibfnamefont {R.~G.}\ \bibnamefont
  {Unanyan}}, \bibinfo {author} {\bibfnamefont {M.}~\bibnamefont
  {Fleischhauer}}, \bibinfo {author} {\bibfnamefont {N.~V.}\ \bibnamefont
  {Vitanov}}, \ and\ \bibinfo {author} {\bibfnamefont {K.}~\bibnamefont
  {Bergmann}},\ }\href@noop {} {\bibfield  {journal} {\bibinfo  {journal}
  {Phys. Rev. A}\ }\textbf {\bibinfo {volume} {66}},\ \bibinfo {pages} {042101}
  (\bibinfo {year} {2002})}\BibitemShut {NoStop}%
\bibitem [{\citenamefont {Kis}\ and\ \citenamefont
  {Paspalakis}(2004)}]{Paspalakis2004}%
  \BibitemOpen
  \bibfield  {author} {\bibinfo {author} {\bibfnamefont {Z.}~\bibnamefont
  {Kis}}\ and\ \bibinfo {author} {\bibfnamefont {E.}~\bibnamefont
  {Paspalakis}},\ }\href@noop {} {\bibfield  {journal} {\bibinfo  {journal} {J.
  Appl. Phys.}\ }\textbf {\bibinfo {volume} {96}},\ \bibinfo {pages} {3435}
  (\bibinfo {year} {2004})}\BibitemShut {NoStop}%
\bibitem [{\citenamefont {Fabian}\ and\ \citenamefont
  {Hohenester}(2005)}]{Fabian2005}%
  \BibitemOpen
  \bibfield  {author} {\bibinfo {author} {\bibfnamefont {J.}~\bibnamefont
  {Fabian}}\ and\ \bibinfo {author} {\bibfnamefont {U.}~\bibnamefont
  {Hohenester}},\ }\href@noop {} {\bibfield  {journal} {\bibinfo  {journal}
  {Phys. Rev. B}\ }\textbf {\bibinfo {volume} {72}},\ \bibinfo {pages}
  {201304(R)} (\bibinfo {year} {2005})}\BibitemShut {NoStop}%
\bibitem [{\citenamefont {Hohenester}\ \emph {et~al.}(2006)\citenamefont
  {Hohenester}, \citenamefont {Fabian},\ and\ \citenamefont
  {Troiani}}]{Troiani2006}%
  \BibitemOpen
  \bibfield  {author} {\bibinfo {author} {\bibfnamefont {U.}~\bibnamefont
  {Hohenester}}, \bibinfo {author} {\bibfnamefont {J.}~\bibnamefont {Fabian}},
  \ and\ \bibinfo {author} {\bibfnamefont {F.}~\bibnamefont {Troiani}},\
  }\href@noop {} {\bibfield  {journal} {\bibinfo  {journal} {Opt. Comm.}\
  }\textbf {\bibinfo {volume} {264}},\ \bibinfo {pages} {426} (\bibinfo {year}
  {2006})}\BibitemShut {NoStop}%
\bibitem [{\citenamefont {Saikin}\ \emph {et~al.}(2008)\citenamefont {Saikin},
  \citenamefont {Emary}, \citenamefont {Steel},\ and\ \citenamefont
  {Sham}}]{Saikin2008}%
  \BibitemOpen
  \bibfield  {author} {\bibinfo {author} {\bibfnamefont {S.~K.}\ \bibnamefont
  {Saikin}}, \bibinfo {author} {\bibfnamefont {C.}~\bibnamefont {Emary}},
  \bibinfo {author} {\bibfnamefont {D.~G.}\ \bibnamefont {Steel}}, \ and\
  \bibinfo {author} {\bibfnamefont {L.~J.}\ \bibnamefont {Sham}},\ }\href@noop
  {} {\bibfield  {journal} {\bibinfo  {journal} {Phys. Rev. B}\ }\textbf
  {\bibinfo {volume} {78}},\ \bibinfo {pages} {235314} (\bibinfo {year}
  {2008})}\BibitemShut {NoStop}%
\bibitem [{\citenamefont {Gauger}\ \emph {et~al.}(2008)\citenamefont {Gauger},
  \citenamefont {Nazir}, \citenamefont {Benjamin}, \citenamefont {Stace},\ and\
  \citenamefont {Lovett}}]{Gauger2008}%
  \BibitemOpen
  \bibfield  {author} {\bibinfo {author} {\bibfnamefont {E.~M.}\ \bibnamefont
  {Gauger}}, \bibinfo {author} {\bibfnamefont {A.}~\bibnamefont {Nazir}},
  \bibinfo {author} {\bibfnamefont {S.~C.}\ \bibnamefont {Benjamin}}, \bibinfo
  {author} {\bibfnamefont {T.~M.}\ \bibnamefont {Stace}}, \ and\ \bibinfo
  {author} {\bibfnamefont {B.~W.}\ \bibnamefont {Lovett}},\ }\href@noop {}
  {\bibfield  {journal} {\bibinfo  {journal} {New J. Phys.}\ }\textbf {\bibinfo
  {volume} {10}},\ \bibinfo {pages} {073016} (\bibinfo {year}
  {2008})}\BibitemShut {NoStop}%
\bibitem [{\citenamefont {Lovett}\ \emph {et~al.}(2003)\citenamefont {Lovett},
  \citenamefont {Reina}, \citenamefont {Nazir},\ and\ \citenamefont
  {Briggs}}]{Lovett2003}%
  \BibitemOpen
  \bibfield  {author} {\bibinfo {author} {\bibfnamefont {B.~W.}\ \bibnamefont
  {Lovett}}, \bibinfo {author} {\bibfnamefont {J.~H.}\ \bibnamefont {Reina}},
  \bibinfo {author} {\bibfnamefont {A.}~\bibnamefont {Nazir}}, \ and\ \bibinfo
  {author} {\bibfnamefont {G.~A.}\ \bibnamefont {Briggs}},\ }\href@noop {}
  {\bibfield  {journal} {\bibinfo  {journal} {Phys. Rev. B}\ }\textbf {\bibinfo
  {volume} {68}},\ \bibinfo {pages} {205319} (\bibinfo {year}
  {2003})}\BibitemShut {NoStop}%
\bibitem [{\citenamefont {Nazir}\ \emph {et~al.}(2005)\citenamefont {Nazir},
  \citenamefont {Lovett}, \citenamefont {Barrett}, \citenamefont {Reina},\ and\
  \citenamefont {Briggs}}]{Lovett2005}%
  \BibitemOpen
  \bibfield  {author} {\bibinfo {author} {\bibfnamefont {A.}~\bibnamefont
  {Nazir}}, \bibinfo {author} {\bibfnamefont {B.~W.}\ \bibnamefont {Lovett}},
  \bibinfo {author} {\bibfnamefont {S.~D.}\ \bibnamefont {Barrett}}, \bibinfo
  {author} {\bibfnamefont {J.~H.}\ \bibnamefont {Reina}}, \ and\ \bibinfo
  {author} {\bibfnamefont {G.~A.}\ \bibnamefont {Briggs}},\ }\href@noop {}
  {\bibfield  {journal} {\bibinfo  {journal} {Phys. Rev. B}\ }\textbf {\bibinfo
  {volume} {71}},\ \bibinfo {pages} {045334} (\bibinfo {year}
  {2005})}\BibitemShut {NoStop}%
\bibitem [{\citenamefont {Gywat}\ \emph {et~al.}(2006)\citenamefont {Gywat},
  \citenamefont {Meier}, \citenamefont {Loss},\ and\ \citenamefont
  {Awschalom}}]{Loss2006}%
  \BibitemOpen
  \bibfield  {author} {\bibinfo {author} {\bibfnamefont {O.}~\bibnamefont
  {Gywat}}, \bibinfo {author} {\bibfnamefont {F.}~\bibnamefont {Meier}},
  \bibinfo {author} {\bibfnamefont {D.}~\bibnamefont {Loss}}, \ and\ \bibinfo
  {author} {\bibfnamefont {D.~D.}\ \bibnamefont {Awschalom}},\ }\href@noop {}
  {\bibfield  {journal} {\bibinfo  {journal} {Phys. Rev. B}\ }\textbf {\bibinfo
  {volume} {73}},\ \bibinfo {pages} {125336} (\bibinfo {year}
  {2006})}\BibitemShut {NoStop}%
\bibitem [{\citenamefont {Hartmann}\ \emph {et~al.}(2008)\citenamefont
  {Hartmann}, \citenamefont {Brandao},\ and\ \citenamefont
  {Plenio}}]{Hartmann2008}%
  \BibitemOpen
  \bibfield  {author} {\bibinfo {author} {\bibfnamefont {M.}~\bibnamefont
  {Hartmann}}, \bibinfo {author} {\bibfnamefont {F.~G.}\ \bibnamefont
  {Brandao}}, \ and\ \bibinfo {author} {\bibfnamefont {M.~B.}\ \bibnamefont
  {Plenio}},\ }\href@noop {} {\bibfield  {journal} {\bibinfo  {journal} {Laser
  $\&$ Photon. Rev.}\ }\textbf {\bibinfo {volume} {2}},\ \bibinfo {pages} {527}
  (\bibinfo {year} {2008})}\BibitemShut {NoStop}%
\bibitem [{\citenamefont {Majer}\ \emph {et~al.}(2007)\citenamefont {Majer},
  \citenamefont {Chow}, \citenamefont {Gambetta}, \citenamefont {Koch},
  \citenamefont {Johnson}, \citenamefont {Schreier}, \citenamefont {Frunzio},
  \citenamefont {Schuster}, \citenamefont {Houck}, \citenamefont {Wallraff},
  \citenamefont {Devoret}, \citenamefont {Girvin},\ and\ \citenamefont
  {Schoelkopf}}]{Wallraff2007}%
  \BibitemOpen
  \bibfield  {author} {\bibinfo {author} {\bibfnamefont {J.}~\bibnamefont
  {Majer}}, \bibinfo {author} {\bibfnamefont {J.~M.}\ \bibnamefont {Chow}},
  \bibinfo {author} {\bibfnamefont {J.~M.}\ \bibnamefont {Gambetta}}, \bibinfo
  {author} {\bibfnamefont {J.}~\bibnamefont {Koch}}, \bibinfo {author}
  {\bibfnamefont {B.~R.}\ \bibnamefont {Johnson}}, \bibinfo {author}
  {\bibfnamefont {J.~A.}\ \bibnamefont {Schreier}}, \bibinfo {author}
  {\bibfnamefont {L.}~\bibnamefont {Frunzio}}, \bibinfo {author} {\bibfnamefont
  {D.~I.}\ \bibnamefont {Schuster}}, \bibinfo {author} {\bibfnamefont {A.~A.}\
  \bibnamefont {Houck}}, \bibinfo {author} {\bibfnamefont {A.}~\bibnamefont
  {Wallraff}}, \bibinfo {author} {\bibfnamefont {M.~H.}\ \bibnamefont
  {Devoret}}, \bibinfo {author} {\bibfnamefont {S.~M.}\ \bibnamefont {Girvin}},
  \ and\ \bibinfo {author} {\bibfnamefont {R.~J.}\ \bibnamefont {Schoelkopf}},\
  }\href@noop {} {\bibfield  {journal} {\bibinfo  {journal} {Nature}\ }\textbf
  {\bibinfo {volume} {449}},\ \bibinfo {pages} {443} (\bibinfo {year}
  {2007})}\BibitemShut {NoStop}%
\bibitem [{\citenamefont {Filipp}\ \emph {et~al.}(2011)\citenamefont {Filipp},
  \citenamefont {Goppl}, \citenamefont {Fink}, \citenamefont {Baur},
  \citenamefont {Bianchetti}, \citenamefont {Steffen},\ and\ \citenamefont
  {Wallraff}}]{Wallraff2011}%
  \BibitemOpen
  \bibfield  {author} {\bibinfo {author} {\bibfnamefont {S.}~\bibnamefont
  {Filipp}}, \bibinfo {author} {\bibfnamefont {M.}~\bibnamefont {Goppl}},
  \bibinfo {author} {\bibfnamefont {J.~M.}\ \bibnamefont {Fink}}, \bibinfo
  {author} {\bibfnamefont {M.}~\bibnamefont {Baur}}, \bibinfo {author}
  {\bibfnamefont {R.}~\bibnamefont {Bianchetti}}, \bibinfo {author}
  {\bibfnamefont {L.}~\bibnamefont {Steffen}}, \ and\ \bibinfo {author}
  {\bibfnamefont {A.}~\bibnamefont {Wallraff}},\ }\href@noop {} {\bibfield
  {journal} {\bibinfo  {journal} {Phys. Rev. A}\ }\textbf {\bibinfo {volume}
  {83}},\ \bibinfo {pages} {063827} (\bibinfo {year} {2011})}\BibitemShut
  {NoStop}%
\bibitem [{\citenamefont {Morley}\ \emph {et~al.}(2010)\citenamefont {Morley},
  \citenamefont {Warner}, \citenamefont {Stoneham}, \citenamefont {Greenland},
  \citenamefont {van Tol}, \citenamefont {Kay},\ and\ \citenamefont
  {Aeppli}}]{Morley2010}%
  \BibitemOpen
  \bibfield  {author} {\bibinfo {author} {\bibfnamefont {G.~W.}\ \bibnamefont
  {Morley}}, \bibinfo {author} {\bibfnamefont {M.}~\bibnamefont {Warner}},
  \bibinfo {author} {\bibfnamefont {A.~M.}\ \bibnamefont {Stoneham}}, \bibinfo
  {author} {\bibfnamefont {P.~T.}\ \bibnamefont {Greenland}}, \bibinfo {author}
  {\bibfnamefont {J.}~\bibnamefont {van Tol}}, \bibinfo {author} {\bibfnamefont
  {C.~W.~M.}\ \bibnamefont {Kay}}, \ and\ \bibinfo {author} {\bibfnamefont
  {G.}~\bibnamefont {Aeppli}},\ }\href@noop {} {\bibfield  {journal} {\bibinfo
  {journal} {Nature Materials}\ }\textbf {\bibinfo {volume} {9}},\ \bibinfo
  {pages} {725} (\bibinfo {year} {2010})}\BibitemShut {NoStop}%
\bibitem [{\citenamefont {Quiroga}\ and\ \citenamefont
  {Johnson}(1999)}]{Quiroga99}%
  \BibitemOpen
  \bibfield  {author} {\bibinfo {author} {\bibfnamefont {L.}~\bibnamefont
  {Quiroga}}\ and\ \bibinfo {author} {\bibfnamefont {N.~F.}\ \bibnamefont
  {Johnson}},\ }\href@noop {} {\bibfield  {journal} {\bibinfo  {journal} {Phys.
  Rev. Lett.}\ }\textbf {\bibinfo {volume} {83}},\ \bibinfo {pages} {2270}
  (\bibinfo {year} {1999})}\BibitemShut {NoStop}%
\bibitem [{\citenamefont {Obada}\ and\ \citenamefont
  {Abdel-Aty}(2007)}]{Obada2007}%
  \BibitemOpen
  \bibfield  {author} {\bibinfo {author} {\bibfnamefont {A.~S.~-F.}\
  \bibnamefont {Obada}}\ and\ \bibinfo {author} {\bibfnamefont
  {M.}~\bibnamefont {Abdel-Aty}},\ }\href@noop {} {\bibfield  {journal}
  {\bibinfo  {journal} {Phys. Rev. B}\ }\textbf {\bibinfo {volume} {75}},\
  \bibinfo {pages} {195310} (\bibinfo {year} {2007})}\BibitemShut {NoStop}%
\bibitem [{\citenamefont {Landau}(1932)}]{Landau}%
  \BibitemOpen
  \bibfield  {author} {\bibinfo {author} {\bibfnamefont {L.}~\bibnamefont
  {Landau}},\ }\href@noop {} {\bibfield  {journal} {\bibinfo  {journal} {Phys.
  Sov. Union}\ }\textbf {\bibinfo {volume} {2}},\ \bibinfo {pages} {46}
  (\bibinfo {year} {1932})}\BibitemShut {NoStop}%
\bibitem [{\citenamefont {Zener}(1932)}]{Zener}%
  \BibitemOpen
  \bibfield  {author} {\bibinfo {author} {\bibfnamefont {C.}~\bibnamefont
  {Zener}},\ }\href@noop {} {\bibfield  {journal} {\bibinfo  {journal} {Proc.
  R. Soc. London}\ }\textbf {\bibinfo {volume} {137}},\ \bibinfo {pages} {696}
  (\bibinfo {year} {1932})}\BibitemShut {NoStop}%
\bibitem [{\citenamefont {Yamauchi}\ \emph {et~al.}(2005)\citenamefont
  {Yamauchi}, \citenamefont {Komori}, \citenamefont {Morohashi}, \citenamefont
  {Goshima}, \citenamefont {Sugaya},\ and\ \citenamefont
  {Takagahara}}]{YamauchiAPL05}%
  \BibitemOpen
  \bibfield  {author} {\bibinfo {author} {\bibfnamefont {S.}~\bibnamefont
  {Yamauchi}}, \bibinfo {author} {\bibfnamefont {K.}~\bibnamefont {Komori}},
  \bibinfo {author} {\bibfnamefont {I.}~\bibnamefont {Morohashi}}, \bibinfo
  {author} {\bibfnamefont {K.}~\bibnamefont {Goshima}}, \bibinfo {author}
  {\bibfnamefont {T.}~\bibnamefont {Sugaya}}, \ and\ \bibinfo {author}
  {\bibfnamefont {T.}~\bibnamefont {Takagahara}},\ }\href@noop {} {\bibfield
  {journal} {\bibinfo  {journal} {Appl. Phys. Lett.}\ }\textbf {\bibinfo
  {volume} {87}},\ \bibinfo {pages} {182103} (\bibinfo {year}
  {2005})}\BibitemShut {NoStop}%
\bibitem [{\citenamefont {Krenner}\ \emph {et~al.}(2005)\citenamefont
  {Krenner}, \citenamefont {Sabathil}, \citenamefont {Clark}, \citenamefont
  {Kress}, \citenamefont {Schuh}, \citenamefont {Bichler}, \citenamefont
  {Abstreiter},\ and\ \citenamefont {Finley}}]{KrennerPRL05}%
  \BibitemOpen
  \bibfield  {author} {\bibinfo {author} {\bibfnamefont {H.~J.}\ \bibnamefont
  {Krenner}}, \bibinfo {author} {\bibfnamefont {M.}~\bibnamefont {Sabathil}},
  \bibinfo {author} {\bibfnamefont {E.~C.}\ \bibnamefont {Clark}}, \bibinfo
  {author} {\bibfnamefont {A.}~\bibnamefont {Kress}}, \bibinfo {author}
  {\bibfnamefont {D.}~\bibnamefont {Schuh}}, \bibinfo {author} {\bibfnamefont
  {M.}~\bibnamefont {Bichler}}, \bibinfo {author} {\bibfnamefont
  {G.}~\bibnamefont {Abstreiter}}, \ and\ \bibinfo {author} {\bibfnamefont
  {J.~J.}\ \bibnamefont {Finley}},\ }\href@noop {} {\bibfield  {journal}
  {\bibinfo  {journal} {Phys. Rev. Lett.}\ }\textbf {\bibinfo {volume} {94}},\
  \bibinfo {pages} {057402} (\bibinfo {year} {2005})}\BibitemShut {NoStop}%
\bibitem [{\citenamefont {Hill}\ and\ \citenamefont
  {Wootters}(1997)}]{Wootters97}%
  \BibitemOpen
  \bibfield  {author} {\bibinfo {author} {\bibfnamefont {S.}~\bibnamefont
  {Hill}}\ and\ \bibinfo {author} {\bibfnamefont {W.~K.}\ \bibnamefont
  {Wootters}},\ }\href@noop {} {\bibfield  {journal} {\bibinfo  {journal}
  {Phys. Rev. Lett.}\ }\textbf {\bibinfo {volume} {78}},\ \bibinfo {pages}
  {5022} (\bibinfo {year} {1997})}\BibitemShut {NoStop}%
\end{thebibliography}

\end{document}